\documentclass[journal]{vgtc}                     

\onlineid{1245}
\usepackage{amsmath} 

\vgtccategory{Research}

\vgtcpapertype{application/design study}

\title{Visual Exploration of Stopword Probabilities in Topic Models}

\author{%
  Shuangjiang Xue,
  \authororcid{Pierre Le Bras}{0000-0002-0670-6552},
  \authororcid{David A. Robb}{0000-0003-4514-959X},
  \authororcid{Mike J. Chantler}{0000-0002-8381-1751}, and
  \authororcid{Stefano Padilla}{0000-0002-7104-8349}
}

\authorfooter{
  \item
  	Authors are with Heriot-Watt University.
  	E-mail: \{sx6, Pierre.Le\_Bras, D.A.Robb, M.J.Chantler, S.Padilla\}@hw.ac.uk.
}

\abstract{%
  Stopword removal is a critical stage in many Machine Learning methods but often receives little consideration, it interferes with the model visualizations and disrupts user confidence. Inappropriately chosen or hastily omitted stopwords not only lead to suboptimal performance but also significantly affect the quality of models, thus reducing the willingness of practitioners and stakeholders to rely on the output visualizations. This paper proposes a novel extraction method that provides a corpus-specific probabilistic estimation of stopword likelihood and an interactive visualization system to support their analysis. We evaluated our approach and interface using real-world data, a commonly used Machine Learning method (Topic Modelling), and a comprehensive qualitative experiment probing user confidence. The results of our work show that our system increases user confidence in the credibility of topic models by (1) returning reasonable probabilities, (2) generating an appropriate and representative extension of common stopword lists, and (3) providing an adjustable threshold for estimating and analyzing stopwords visually. Finally, we discuss insights, recommendations, and best practices to support practitioners while improving the output of Machine Learning methods and topic model visualizations with robust stopword analysis and removal.
}


\teaser{
  \centering
  \includegraphics[width=\linewidth-1pt, alt={Our stopword analysis interface, including a topic model representation on the left, and word inspection panels on the right.}]{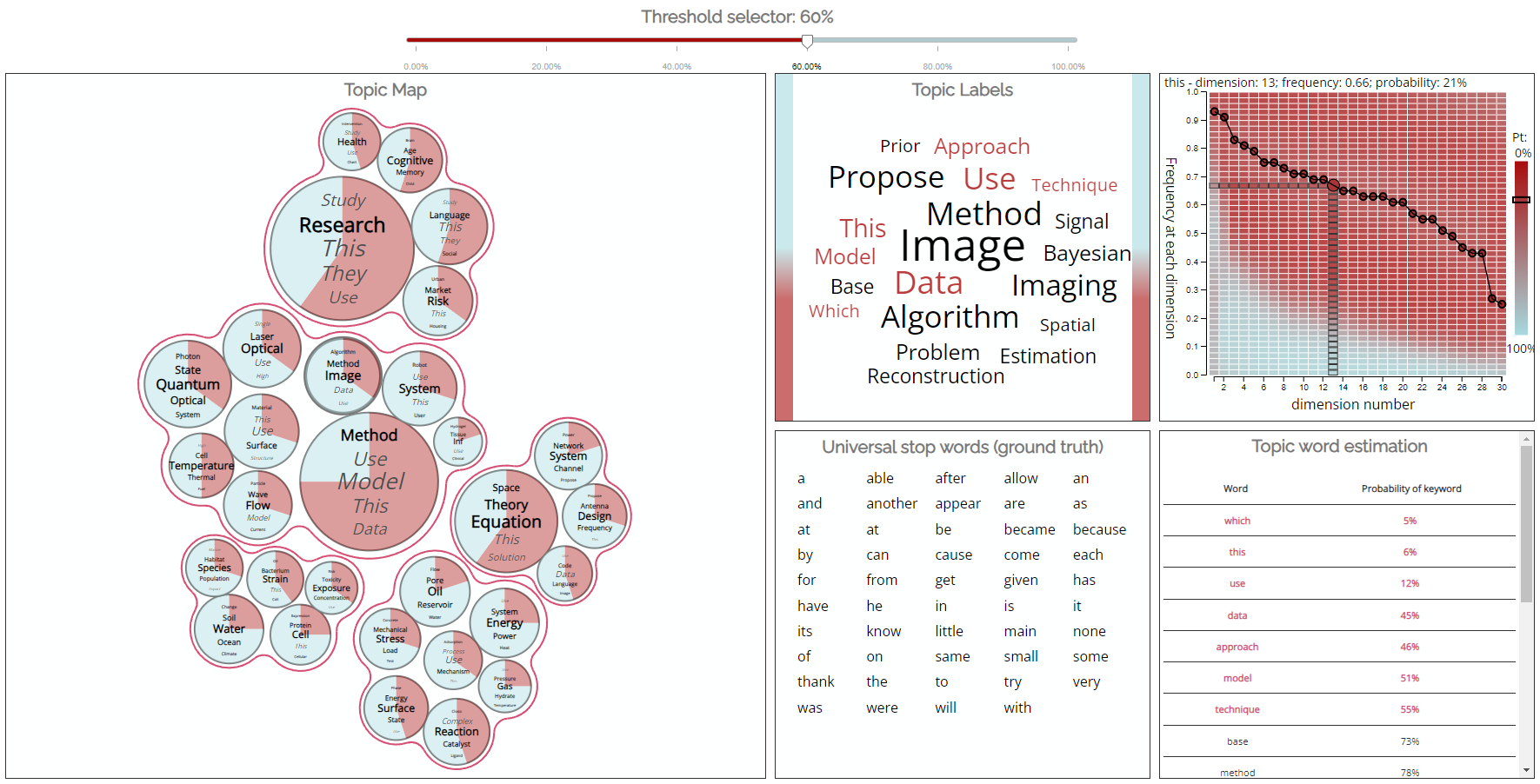}
  \caption{%
        Our stopword analysis visualization interface. We combine the topic model visualization system from Le Bras et al. \cite{le2020visualising} with a novel 2D visualization of an approximate Gaussian Process Classification model: the GPC Matrix (top right-hand side). Users get an overview of the clustered topics on the left-hand side panel. Upon selecting a topic, a wordcloud displays the top words. Selecting a word highlights its probability of being a topic word on the GPC Matrix. On the top, users can adjust a threshold to highlight stopwords with different probability levels. Colors were adjusted to be photocopy-safe.%
  }
  \label{fig:interface}
}

\graphicspath{{figs/}{figures/}{pictures/}{images/}{./}} 

\usepackage{tabu}                      
\usepackage{booktabs}                  
\usepackage{lipsum}                    
\usepackage{mwe}                       

\usepackage{mathptmx}                  

\usepackage[none]{hyphenat}
\usepackage{dingbat}

\begin{document}



\maketitle

 \section{Introduction} 

Like many data cleaning and pre-processing steps, stopword removal is a crucial stage in text-based Machine Learning (ML) algorithms. In topic modelling applications, for example, it enhances overall performance and provides a significant improvement in the generation of meaningful and interpretable topics \cite{stopwordRethinking}. However, this stopword removal exercise is not without challenges for practitioners aiming to present their analysis confidently. 
For visualization tasks in particular, it stands out as a crucial step for enhancing the quality of the final outcomes provided to end users. From our experience, we have identified three key challenges: (a) establishing a threshold under which a word can be considered a stopword; (b) accounting for corpus-specific features; and (c) rationalizing the choice of stopwords to stakeholders.

Stopwords are defined as words with a high frequency but little or no meaningful information \cite{2020stopwordIr}. For topic modelling tasks that involve analyzing large document corpora, words that do not contribute to the expressiveness of topics are typically considered stopwords \cite{boyd2014care}. Along universal stopwords (e.g., ``\textit{a}'', ``\textit{the}'' or ``\textit{and}'') there are also corpus-specific stopwords \cite{corpusSpecStopwords}: for example, in a 2020 medical corpus \cite{wang-etal-2020-cord}, the word ``\textit{coronavirus}'' would appear in almost every document. Typical visual topic model analysis systems (e.g., LDAVis \cite{sievert2014ldavis} or BERTopic \cite{bert}) integrate universal stopword removal and a visual inspection of the remaining words' statistics. 

Detecting corpus-specific stopwords is a manual and laborious process with little support for interpretability. Topic model processes usually provide scores that denote the importance of words within a topic, a document or the corpus. However, users are often required to go through words one by one without seeing overall distribution patterns that would highlight the impact of removing specific stopwords.


This research assumes that the latent distribution patterns of a few universal stopwords, generally agreed upon by users, can be the basis for classifying -- and removing -- every other stopword in any typical corpus, whether common or domain-specific. This approach reduces the variability in stopword removal across different domains and provides more consistency in the overall process. This paper introduces a dynamic probabilistic estimation method for stopword removal based on the Gaussian processes classification (GPC) model. A GPC model is a supervised ML model that can learn the latent distribution of input data. Then, the model can provide a likelihood of any word being a corpus stopword (Section~\textbf{\ref{sec:method}~Method}).

On its own, this method can be integrated into automated applications. However, from our previous experience in dealing with topic model tasks [anonymized], users tend to have different opinions as to whether a word is a stopword or not. These are primarily professional judgments based on user expertise, intuition or awareness of the corpus. We consider these judgments valuable when extracting stopwords for different tasks. However, they can cause bias due to unnecessarily removed words \cite{stopwordRethinking} and a possible misunderstanding of the corpus. Beyond an algorithmic detection of stopwords, our work aims to help practitioners make more prudent and evidence-based decisions that minimize bias. Our method design therefore targets stopword removal at different probability levels. Combined with an interactive topic model visualization system (Figure \ref{fig:interface}), we provide a visual probability-driven analysis of stopwords, where users can select a probability threshold and access the influence of their choice in the overall model and individual topics (Section~\textbf{\ref{sec:interface}~Interface}).

By combining a probabilistic estimation method and interactive visual tools, our work aims to improve practitioners' confidence in their analysis and alleviate the aforementioned issues. To assess the impact of our tool on confidence, this paper presents a controlled qualitative user study designed to mimic a real-world scenario (Section~\textbf{\ref{sec:study}~Study}). The study evaluated whether our method provides a correct stopword list that users are willing to implement with confidence. The study was conducted in four phases, consisting of semi-structured interviews with questions related to tasks. Based on our method and experimental results, we answer four research questions and conclude with recommendations for practitioners and users (Section~\textbf{\ref{sec:results}~Results}).

In summary, this work has three major contributions:
\begin{itemize}
    \item We propose \textbf{a novel dynamic stopword extraction method}. Given a universal stopword list, our method estimates the probabilities of every word to be a stopword within a specific topic-modelled corpus.
    \item We introduce \textbf{an interactive visualization interface for the analysis of stopwords} which combines a topic model display along with a 2-dimensional visualization of the approximate GPC model.
    \item We \textbf{explore the influence of our method and visualization system on users' confidence} in a qualitative study. From this, we discuss insights and recommendations to improve and aid future practice.
\end{itemize}

\section{Related Work}

This section describes previous work related to stopword estimation methods and topic model visualization systems. We also review how existing systems visualize stopwords to users.

\subsection{Stopword Estimation}

Despite the many advancements and improvements in topic modelling analysis \cite{Churchill2022}, stopword removal continues to have a crucial impact on the performance of topic models such as BERTopic and LDA. The typical solution prescribed to address this problem consists in filtering out an established static list of common stopwords \cite{manning2014stanford}. For instance, two of the most commonly used topic model visualization systems, LDAVis \cite{sievert2014ldavis} and BERTopic \cite{grootendorst2022bertopic}, offer users the option to remove predefined stopwords without disclosing the stopword list or permitting any augmentation with additional words. Although fixed lists typically contain most stopwords and are available in many languages, works by Boyd-Graber et al. \cite{boyd2014care} and Nothman et al. \cite{nothman2018stop} have established that they are often incomplete. As described by Churchill et al. \cite{churchill2018}, \textit{flood words} or corpus-specific stopwords often exist in corpora, in particular domain-specific ones. These words tend to flood the resulting topics, without meaningfully contributing to their quality. An illustrative instance is the word ``\textit{coronavirus}'' within The COVID-19 Open Research Dataset \cite{wang-etal-2020-cord}.

A typical approach to enhancing these lists (and improving the topic model outputs and visualization) is to manually choose additional words: practitioners and expert users would review the model topic-by-topic, explore their descriptions word-by-word, and highlight words to remove in the next modelling iteration. The review process of all words in every topic is time-consuming. Meanwhile, results lack in reusability across different corpora \cite{gerlach2019universal} and are inadequate due to limited user observation \cite{hao2008automatic}. To address these limitations, various dynamic stopword removal techniques have been proposed. Term evaluation scores such as term-frequency inverse document-frequency (TF-IDF) \cite{joachims1997probabilistic}, corpus TF-IDF (c-TF-IDF) \cite{grootendorst2022bertopic}, term relevance \cite{sievert2014ldavis}, conditional entropy \cite{gerlach2019universal} and saliency \cite{chuang2012termite} have been proposed to evaluate terms in language processing tasks. While TD-IDF quantifies the importance of a word in a document, c-TF-IDF does it for the entire corpus. Term relevance uses a weight $\lambda$ to compare a word's probability in a given topic against its lift \cite{taddy2012} (ratio between the word probability in a topic and the marginal word probability in the corpus). Conditional entropy serves as a specific term relevance measurement, which indicates the degree to which a term is associated with a particular topic. Saliency assesses the exclusivity of a term within a specific topic by analyzing its probability within that topic and across the entire corpus. Dynamic extraction methods for corpus-specific stopword lists have also been proposed. Wallach et al., for instance,  pointed out that setting as asymmetric prior over $\alpha$ and symmetric prior over $\beta$ for LDA, skew words' distributions and tends to gather stopwords in specific topics \cite{wallach2009rethinking}. More recently, Schofield et al. suggested a topic document mutual information-based method \cite{stopwordRethinking}. 

Nevertheless, despite the application of these automated extraction and evaluation methods, stopword extraction tasks still require human intervention. For example, Sarica et al. \cite{sarica2021stopwords} extract stopwords in the technical area by analyzing an intersection among four independent metrics. With the professional efforts provided to a specific area, achieving a perfect reliable list in human evaluation has proved challenging. Simultaneously, we have noted significant individual differences among users in their judgements of ambiguous words (an observation also discerned in our study, see Section~\ref{sec:results}~Results).

\subsection{Topic Model Visualizations}

Within visualization systems especially, stopword analysis (and removal) is an under-supported functionality. Most topic model visualization systems are typically designed for the presentation of topics and their relations, based on various heuristics: for example, geography \cite{choi2018}, time \cite{liu2020mapping}, multi-dimensional projection \cite{grootendorst2022bertopic}, network \cite{abdul2018}, cluster \cite{le2020visualising} or simple lists \cite{chaney2012visualizing}. Then, common interactions would include the selection of a topic, linked with listing the most relevant words and documents for that topic.

After universal stopwords are removed during vectorization, it has been shown that the default stopword list is incomplete and contains controversial words \cite{nothman2018stop}. Yet, few topic model visualization systems incorporate word visualization features to explain these issues and allow users to analyze topics and corpora dynamically. Termite \cite{chuang2012termite} proposes to present a global view of word distributions across topics in a bubble matrix. Another example, LDAVis \cite{sievert2014ldavis}, displays the ratio of term frequency within a selected topic against the entire corpus and changes the size of all topics based on conditional topic distribution given the term. It also includes measures such as relevance and saliency and uses those to alter the ranking of words in topics interactively. While this provides insights into how words effectively contribute to topics, it does not mark stopwords as such and does not communicate the impact of stopwords in the overall model. BERTopic \cite{grootendorst2022bertopic} provides term visualization based on c-TF-IDF scores indicating a term's importance within a document and the corpus. In summary, if a corpus-specific stopword list needs to be selected with these two topic model visualization systems, the exploration process for identifying stopwords is corpus > topic > word > scores > judgement.

Therefore, we identify the following three major gaps in current exploration processes of stopword analysis:
\begin{itemize}
    \item It is difficult to make well-founded judgments from a topic perspective without considering the features of the topics. There are different patterns of word distributions through topics which represent various levels of it being a stopword.
    \item There is a loss of consistency when switching from one topic to another, the users' mental evaluation criteria of the scores and sizes keep changing between topics.
    \item It can be extremely time-consuming since the user needs to make judgments for every word in every topic. For instance, according to our observation, the automated number of topics given by BERTopic for a large corpus is within the thousands, which makes it impossible to go through a rigorous stopword selection
\end{itemize}

\subsection{Summary}

Ultimately, to our knowledge, there has not been an interactive visualization system capable of identifying a list of stopwords, with a user-adjustable metric, that also explains how such a list is generated.

As described above, we believe that human intervention can enhance practitioner and user confidence in stopword analysis and removal. Therefore, we define it as an incomplete task that requires interpretability \cite{doshi2017towards}. This research aims to minimize the difficulty and increase the accuracy of human decisions. To achieve our goal, we evaluate each term with a corpus-specific ML-based probabilistic score to assist manual intervention and implement an interactive interface to allow an efficient one-drag threshold decision.

The crucial features of this research are the limited number of input stopwords, distribution analysis, and interpretability. From many explainable ML models \cite{roscher2020explainable}, we consider the GPC model \cite{rasmussen2004gaussian} as the ideal model for this task. The GPC model can make inferences based on a small input dataset that allows us to minimize the input universal stopword list, learn latent distribution patterns based on the corpus feature and provide probabilistic measurements of uncertain quantification problems.

\section{Method}\label{sec:method}

This section describes the topic model used in this research, along with the training data and frequency metrics used in our method. It concludes with a presentation of the Gaussian processes classification (GPC) used to generate a probabilistic stopword estimation model.

\subsection{Topic Model}


We designed this work to analyze the users' confidence in stopword removal based on a topic model visualization use case. In this work, we use Latent Dirichlet Allocation (LDA) \cite{lda}, an established topic model method. 

Our study used a corpus of 4920 research publication documents from a university in 2016. The qualitative study described later (Section~\ref{sec:study}) simulates experts presenting their analysis based on the visualized topic model, and this corpus was selected based on our participants' expertise and the study scenario. Considering the visual outcome and complexity of qualitative experimental tasks, the number of topics is set to 30. To generate the topic model, we use recommended hyper-parameter settings \cite{boyd2014care}.




\subsection{Training Data}

\begin{table}
  \caption{List of training stopwords, minimized to avoid over-removal.}
  \label{tab:trainingSW}
  \begin{center}
  \begin{tabular}{c c c|c c c}
  \toprule
  \multicolumn{3}{c|}{\scriptsize{\textit{non-selective stopwords}}} & \multicolumn{3}{c}{\scriptsize{\textit{manually chosen stopwords}}} \\
  a & an & and & able & after & allow \\
  are & as & at & another & appear & became \\
  be & by & for & because & cause & come \\
  from & has & he & can & each & given \\
  in & is & it & get & have & know \\
  its & of & on & little & main & none \\
  at & the & to & same & small & some \\
  was & were & will & thank & try & very \\
  with & & & & & \\
  \bottomrule
\end{tabular}
\end{center}
\end{table}

\begin{table}
  \caption{List of training topic words, extracted from topic descriptors (top two words).}
  \label{tab:trainingTW}
  \begin{center}
  \begin{tabular}{c c c c c}
    \toprule
    age & antenna & bacterium & cell & code \\
    cognitive & complex & data & design & energy \\
    equation & exposure & flow & gas & habitat \\
    health & image & inf & language & laser \\
    market & mechanical & method & model & network \\
    oil & optical & pore & pressure & process \\
    protein & quantum & reaction & research & risk \\
    soil & species & state & strain & stress  \\
    study & surface & system & temperature & theory \\
    this & tissue & toxicity & use & water \\ 
    wave\\ 
    \bottomrule
\end{tabular}
\end{center}
\end{table}

The GPC model is a supervised learning model that requires labelled training data, in this study, \textbf{stopwords} and \textbf{topic words}. For training stopwords, to avoid over-removal, we minimise the input universal stopword list. The baseline of the list contains 25 semantically non-selective words \cite{Reuters-RCV1}. To simulate common practice when users add input stopwords and allow a more realistic exploration and analysis, we manually extended 24 common words based on our experience. Table~\ref{tab:trainingSW} lists the full set of training stopwords. For training topic words, the top two relevant words (given by the LDA model) from each topic are selected. Table~\ref{tab:trainingTW} lists the full set of training topic words.

\subsection{Set of Document Frequency}\label{sec:set}

The aim of this research is to estimate a word classification based on its specific distribution within the topic model. Therefore, all topics are taken into account. Given that the model comprises $n$ topics, for each training word, we feed the GPC model with a vector of dimensions $n$. 

\begin{figure}
  \centering
  \includegraphics[width=1\linewidth]{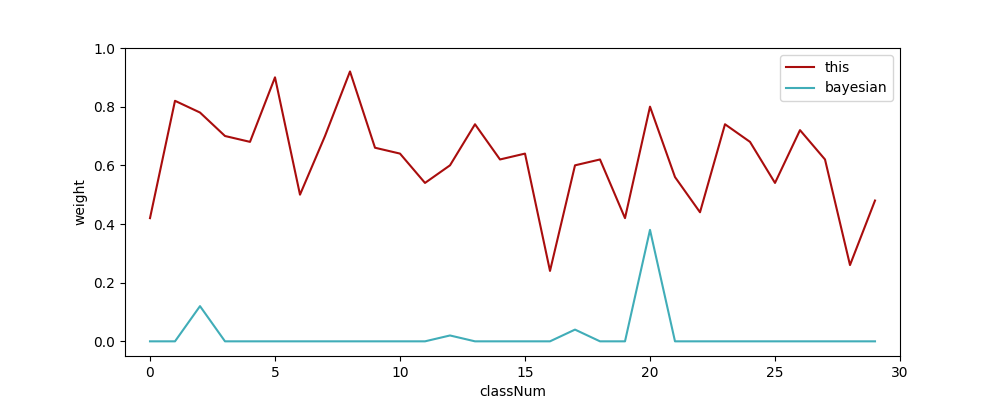}
  \caption{Examples of the $\mathbf{Swdf}_j$ of a representative stopword ``\textit{this}'' and a topic word ``\textit{bayesian}''}
  \label{fig:swtw}
\end{figure}

For each word ${w}_j$, within each topic $i$ ($i= 1,2, \cdots, n$), we set the number of documents in it as $Nd_i$ and the number of documents containing the word $w_j$ as $Nd_iw_j$. The document frequency of the target word $j$ in the topic $i$, $Dfw_{j, i}$, is then calculated as:
\begin{equation}
  Dfw_{j,i} = \frac{Nd_iw_j}{Nd_i}\qquad \mathrm{for}\; i =1,\;2,\cdots,\;n,
\end{equation}
where $0\le Dfw_{j,i} \le 1$. In this work, we set $\mathbf{Swdf}_j=\{Dfw_{j,i}\}_{i=1}^{n}$. For each word $w_j$, the $\mathbf{Swdf}_j$ denotes the distribution over the topics. Figure~\ref{fig:swtw} highlights the difference between  the $\mathbf{Swdf}_j$ of a training stopword and topic word. A pilot experiment proved that the document frequency outperformed the term frequency typically used (see Appendix~\ref{app:pilot}).

\vspace{6 pt}



\begin{figure}
  \includegraphics[width=\linewidth]{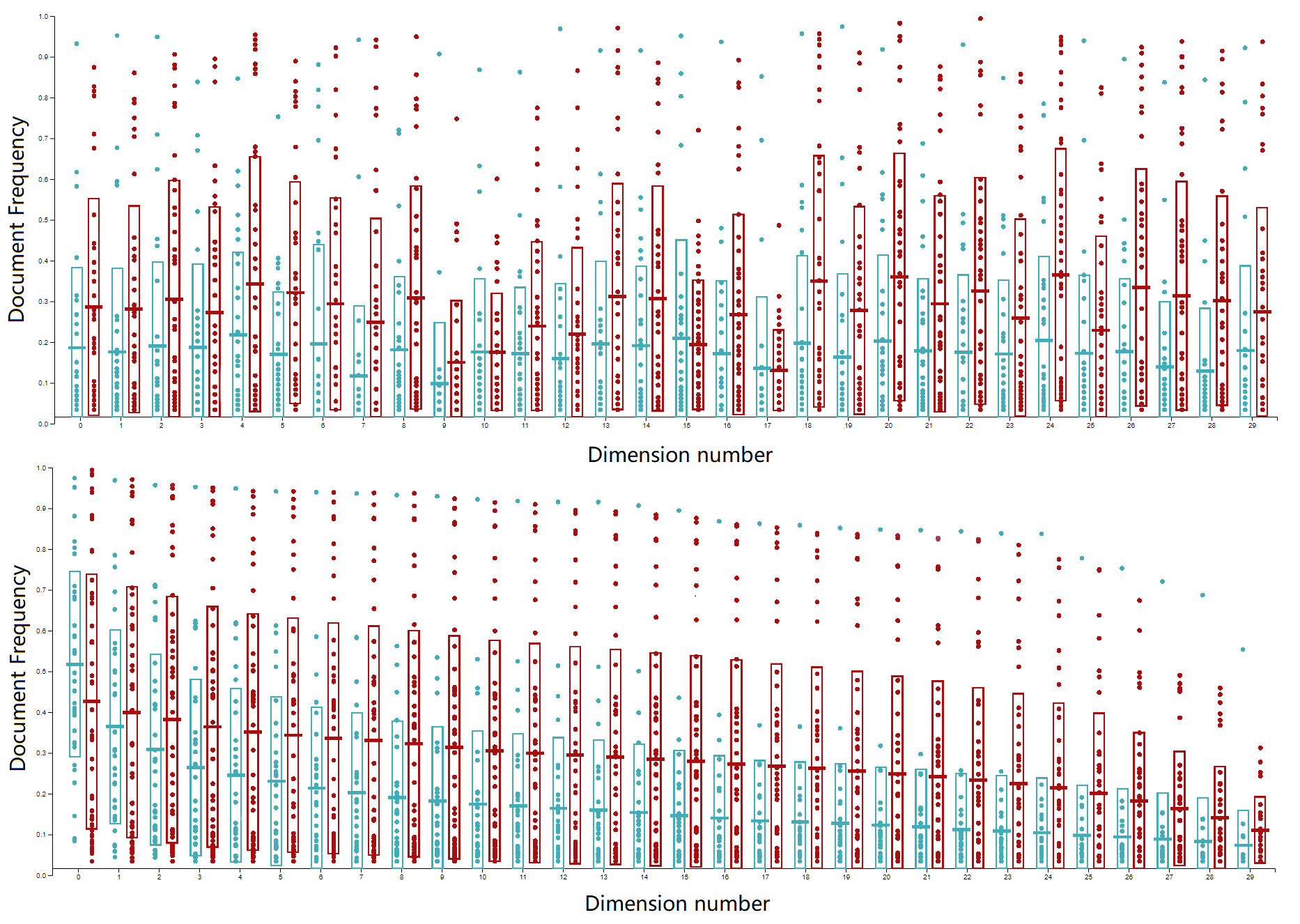}
  \caption{The comparison of the original layout with the monotone decreasing layout. Red stands for stopwords while blue stands for topic words. The rectangular shows 64\% confidence interval of each dimension and the line in the middle stands for the mean.}
  \label{fig:dis-joint}
\end{figure}

For each word $w_j$, our methods sorts the $\mathbf{Swdf}_j$ into monotone decreasing order: $\mathbf{Swdf}_j = \{({Dfw_{j, i})_{max}},\; \cdots,\; ({Dfw_{j, i})_{min}}\}$.
By doing this, we gathered the highest frequencies into particular dimensions since they fall into various topics for different words. In Figure~\ref{fig:dis-joint}, we can easily find that a monotonic decreasing data set (bottom) better distinguishes the two classes. This operation shuffles the order of topics for each word. For example, given two words $w_{j=1}$ and $w_{j=2}$, the topics that achieve the highest document frequency for $w_{1}$ and $w_{2}$ are very likely not the same. Hence we use dimension $h$ instead of topic $i$ when expressing $\mathbf{Swdf}_j=\{Dfw_{j,h}\}_{h=1}^{n}$ to clarify, where ${Dfw_{j,{h=1}} = ({Dfw_{j, i})_{max}}}$, ${Dfw_{j,{h=n}} = ({Dfw_{j, i})_{min}}}$. For a better training outcome, we then normalize the words' $\mathbf{Swdf}$ to make the data set add up to one before training the GPC model by defining:
\begin{equation}
\mathbf{Swdf}_j=\{\frac{Dfw_{j,h}}{\sum_{h=1}^{n}Dfw_{j,h}}\}_{h=1}^{n}.
\label{eq_swdf_resacle}   
\end{equation}

\subsection{Probability Classification - GPC}\label{training}

Since the topic word selection process can be considered as a typical binary ($C = 2$) probabilistic classification task, this work applies a Gaussian processes classification (GPC) model. We used the version implemented by Pedregosa et al. \cite{scikit-learn}. The Gaussian processes and the test prediction probability took the form of class probabilities.

We use a training set $D$ of 100 observations with 51 topic words and 49 stopwords shown in the Table \ref{tab:trainingSW} and \ref{tab:trainingTW}. $D = \{(\mathbf{Swdf}_m, y_m) | m = 1,\;\cdots,\; 100\}$, where $\mathbf{Swdf}_m$ is the input vector of dimension 30 (follows the number of topic), detailed in Equation (\ref{eq_swdf_resacle}). $y$ denotes the class: stopword as ${y=0}$; topic word as ${y=1}$. The GPC model is trained to make inferences about the relationship between inputs $\mathbf{Swdf}$ and classes $\mathbf{y}$. GPC models the class-conditional densities with Gaussians: $p(\mathbf{Swdf}|Y_y) = N(\mu_y,\Sigma_y)$.

Given a latent function $f$ and a test case $\pi(\mathbf{Swdf}_{\ast})$. Following Bayes theorem, the posterior over latent variables, $p(\mathbf{f}|\mathbf{Swdf},\mathbf{y})$, can be expressed as:
\begin{equation}
p(\mathbf{f}|\mathbf{Swdf})=\frac{p(\mathbf{y}|\mathbf{f})p(\mathbf{f}|\mathbf{Swdf})}{p(\mathbf{y}|\mathbf{Swdf})}.
\end{equation}
Thus the model provides the probabilistic prediction of whether a word is a stop word by using its distribution over the latent $f_{\ast}$:
\begin{equation}
    \begin{split}
        \bar\pi_{\ast}\triangle p(y_{\ast}=1|\mathbf{Swdf},\mathbf{y},\mathbf{Swdf}_{\ast})= \int \sigma(f_{\ast})p(f_{\ast}|\mathbf{Swdf},\mathbf{y},\mathbf{Swdf}_{\ast})df_{\ast}.
    \end{split}
\end{equation}

Generally, a Gaussian Process prior combined with a Gaussian likelihood gives rise to a posterior Gaussian process over functions. However, since the targets in classification models are discrete class labels, the Gaussian likelihood is inappropriate \cite{rasmussen2004gaussian}. The posterior process is approximated by a Gaussian process and estimated using Laplace approximation.

After the GPC is trained, for any word within the topic model, the model is able to estimate the probability of topic word, $p_t$, given $\mathbf{Swdf}_{\ast}$ of the word where $p_t=p(y_{\ast}=1|\mathbf{Swdf},\mathbf{y},\mathbf{Swdf}_{\ast})$, and a probability of stopword $p_s=1-p_t$. With these probabilities established, a list of stopwords can then be extracted, using a threshold $p_t$. For example, Table~\ref{tab:sw60} shows the list of stopwords with $p_t < 60\%$.





\begin{table}
  \caption{The full extracted stopwords list with $p_t$ lower than $60\%$. Words in bold are typically not included in common stopword lists, however, they correspond to stopwords one would expect in a corpus of university publications (words too generic in this corpus to carry significant semantic meaning).}
  \label{tab:sw60}
  \begin{center}
  \begin{tabular}{c c c c c c}
    \toprule
    all    & also    & \textbf{analysis}  & \textbf{approach}  & between    & but   \\
    \textbf{data}   & during  & here      & \textbf{high}      & into       & \textbf{model} \\
    more   & not     & over      & \textbf{paper}     & \textbf{potential}  & \textbf{show}  \\
    \textbf{study}  & such    & these     & this      & through    & \textbf{time}  \\
    two    & \textbf{use}     & which \\
    \bottomrule
\end{tabular}
\end{center}
\end{table}

\section{Interface}\label{sec:interface}

In addition to the GPC model, and to help users interpret it, we have designed an interactive visualization interface providing the following elements: 
\begin{itemize}
    \item An interactive stopwords threshold selector based on model estimates
    \item How selected stopwords are distributed across the topics
    \item The model estimation of topic words and stopwords
    \item A visual explanation of the principle behind stopword estimation
\end{itemize}

Our interface, shown in Figure~\ref{fig:interface}), is based on the work of Le Bras et al. \cite{le2020visualising}. This system presents two advantages. First, the general interface structure permits users to get an overview of the model, along with detailed topic information on demand \cite{Shneiderman2003}. Second, its modular design allows the integration of novel visualization blocks. We use two complementary colours in the interface representing different types of probabilities: red for stopwords and blue for topic words (note that colors were adjusted to be photocopy-safe).

\subsection{Threshold Selection}

As described in section \ref{sec:method}~Method, the GPC model gives every word a probabilistic estimation, $Ps$ and $Pt$. We can always divide words into two classes by setting a threshold for $Pt$(or $Ps$). In our pilot experiments (details in Appendix~\ref{app:pilot}), we found that the dividing edge changes through different corpora and input words. Moreover, as shown in our user experiment, the tolerance for what is or isn't a stopword varies between users. Hence, \textbf{there is no ideal threshold value}. Therefore, we introduced a selector at the top of the interface. Users can set the threshold of $Pt$ by positioning the handle between the left-hand side($0\%$, all words are topic words) and the right-hand side($100\%$, all words are stopwords). Coordinated with other views on the interface, this selector allows users to \textbf{extract stopwords with a one-drag operation}. Meanwhile, the driven changes in other panels help users acknowledge the impact of their removal on the whole topic model.

\begin{figure*}[ht]
  \centering
  \begin{subfigure}{0.33\linewidth}
  	\centering
  	\includegraphics[width=1\linewidth, alt={GPC Matrix for the word 'this', a word that should be considered a stopword. Connected circles are only in the high stopword probability cells (red).}]{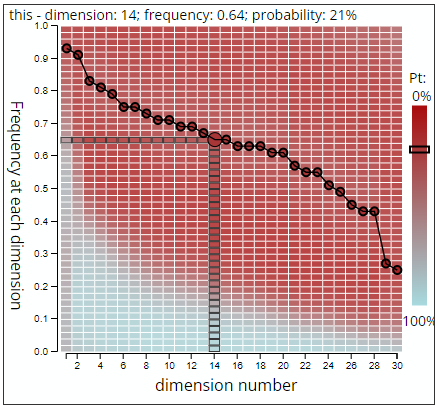}
  	\caption{Typical stopword: ``\textit{this}''.}
  	\label{fig:2dgp_stop}
  \end{subfigure}%
  \hfill%
  \begin{subfigure}{0.33\linewidth}
  	\centering
  	\includegraphics[width=1\linewidth, alt={GPC Matrix for the word 'data', a word that could be considered a stopword. Connected circles are on the edge between the high topic-word probability cells (blue) and high stopword probability cells (red).}]{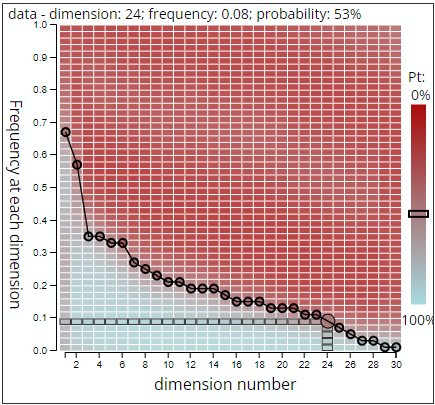}
  	\caption{Borderline stopword/topic word: ``\textit{data}''.}
  	\label{fig:2dgp_middle}
  \end{subfigure}%
  \hfill%
  \begin{subfigure}{0.33\linewidth}
  	\centering
  	\includegraphics[width=1\linewidth, alt={GPC Matrix for the word 'bayesian', a word that should not be considered a stopword. Connected circles are only in the high topic word probability cells (blue).}]{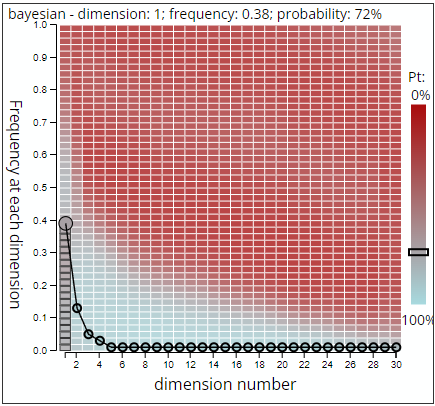}
  	\caption{Typical topic word: ``\textit{bayesian}''.}
  	\label{fig:2dgp_topic}
  \end{subfigure}%
  \subfigsCaption{The \textbf{GPC Matrix}: a 2-D approximate visualization of the GPC model.  Cells divide the problem space into a $30 \times 50$ grid. The background colour of each cell corresponds to the trained probabilities of a word being a topic word, given a dimension ($h$) and a document frequency ($df$). A blue background signifies a high probability of being a topic word. A red background signifies a high probability of being a stopword. Upon selecting a word (within a topic), the user is provided with the superimposition of the estimated probabilities for that word as connected black circles. The three examples above provide the typical three cases: \subref{fig:2dgp_stop} a clear stopword, \subref{fig:2dgp_middle} a borderline word and \subref{fig:2dgp_topic} a clear topic word.}
  \label{fig:2dgp}
\end{figure*}

\subsection{Topic Map}

This panel visualizes the topic model (shown on the left side of Figure \ref{fig:interface}). Thirty circles represent the thirty topics, with their positions based on similarities. The words within the circles are the top-related words for each topic. In this work, we added two sectors to indicate the ratios between topic words and stopwords. This ratio was computed from the top-20 words for each topic. The sectors allow users to get insights into how their choice of threshold influences the classification of stopwords in each topic.


\subsection{GPC Matrix}

Visualizing how a 30-dimensional model classifies words is an intricate problem. Instead, we have designed a 2-D approximate visualization: the GPC Matrix. This approximation is calculated by changing the format of the model's input data. As described in section \ref{sec:method}~Method, we use the same set of training words. However, instead of the 30-dimensional input vectors, each word input is a set of thirty 2-dimensional vectors $\{(Dfw_{j,h},\;h,\;y_j)\}_{h=1}^{30}$, where $Dfw_{j,h}$ is the document frequency for the word $w_j$ with the same label $y_j$ across $h$ dimensions. Therefore, the new training set for 50 words becomes the set of $\{(Dfw_{m,h},\;h,\;y_m)\}_{h=1}^{30}$ for $m=1,\;\cdots,\;50$. This 2-D model partially loses co-relations between dimensions in the covariance matrix, in return however, it lets us present how the GPC model makes decisions to users.

To visualize the GPC model, we establish a matrix with the two input factors, where the horizontal axis is the dimension $h$, and the vertical axis is the document frequency $Dfw$. There are 30 dimensions (topics) in our model. For $Dfw$, we evenly discretize its range ($\left[0,\; 1\right]$) into 50 points, hence the vertical axis unit here is 0.02\footnote{Other units may used.}. We now have a $30 \times 50$ problem (probability) space from which the 2D GPC model can predict a probability of topic word ($p_{t_h}$) on dimension $h$ after training. We represent these probabilities with different colors for each cell in our matrix: we denote $p_t=1$ in blue and $p_t=0$ in red. Thus we can display the model as the background shown in Figure \ref{fig:2dgp}. These background colours showed the looks of the 2-D GPC model through the two-dimensional problem space. A legend of the probability-colour transformation is displayed on the right of the matrix.

Figure \ref{fig:2dgp_stop} shows the GPC Matrix when the word ``\textit{this}'' is selected from the topic label panel. The black circles denote the position of document frequencies on each dimension, and the solid lines that connect the circles represent how $p_t$ of the word ``\textit{this}'' distributes through the problem space. The colour in each black circle indicates how likely/unlikely the 2-D GPC model considers the word to be a topic word on the given dimension. The 30-D GPC making a joint estimation means that redder circles indicate a higher probability of a word being a stopword overall.

In contrast, Figure \ref{fig:2dgp_topic} shows the GPC Matrix when a word like ``\textit{bayesian}'' is selected. The probability line is now overlapping a region of the probability space with higher probabilities of topic words. Figure \ref{fig:2dgp_middle} shows the GPC Matrix for a word like ``\textit{data}'', one sitting on the edge between stopword or topic word.

This panel is designed to improve user confidence when the model predictions are different from the user's expectations. People tend to make a judgement of stop/topic words based on their own experience or expertise. However, it is usually hard to grasp a corpus's overall situation, and experiences are sometimes not comprehensive. Our visualization provides objective and non-empirical information, which could be a good reference for the user to make a judgement.

\subsection{Helper Panels}

The three other helper panels in the interface shown in Figure \ref{fig:interface} provide the following functionalities: (a) \textbf{Topic labels} -- a topic word cloud shown when users select a bubble from the topic map; stopwords extracted by the threshold are highlighted in red, and two rectangles on both sides indicate the ratio of stopwords under the threshold; (b) \textbf{Universal stopwords} -- the ``ground truth'' list of training stopwords showing users the benchmark of removal; (c) \textbf{Topic word estimation} -- the list of words and their estimate $p_t$ in the selected topic (low to high), indicating the stop/topic word split with coloured words (red if $p_t$ is lower than the threshold).

\newcommand{\RQOn}{\textbf{\textcolor{cyan!70!black}{RQ1}}}
\newcommand{\RQTw}{\textbf{\textcolor{green!70!black}{RQ2}}}
\newcommand{\RQTh}{\textbf{\textcolor{orange!70!black}{RQ3}}}
\newcommand{\RQFo}{\textbf{\textcolor{violet!70!black}{RQ4}}}

\begin{table*}[ht]
  \caption{The 4 stopword lists used in our study.}
  \label{tab:lists}
  \renewcommand{\arraystretch}{1.5} 
  \centering%
  \begin{tabu}{ >{\centering\arraybackslash}p{0.04\linewidth} p{0.44\linewidth} p{0.44\linewidth} }
    \toprule
    \textbf{Name} & \textbf{Description} & \textbf{Purpose} \\
    \midrule
    \textbf{L1} & List of common stopwords \cite{manning2014stanford} manually augmented with corpus-specific stopwords for academic publications. 554 stopwords in total.  & Provide a typical curated stopword list used in topic modelling. \\
    \textbf{L2} & List of 49 common stopwords generally agreed by practitioners, used to train the GPC models (Section~\ref{training}). & Compare against L1 to gather preferences between long curated lists and short agreed lists. (\RQOn) \\
    \textbf{L3} & List of 27 corpus-specific stopwords extracted using the GPC model, with a $p_t$ lower than $60\%$ (Table~\ref{tab:sw60}). & Compare L2+L3 against L1 to gather preferences and confidence between manually curated and automated lists. (\RQOn, \RQFo) \\
    \textbf{L4} & List of 23 corpus-specific stopwords extracted using the 2-D approximate GPC model. & Compare against L3 to gather preferences and understanding of the GPC Matrix. (\RQTw) \\
    \bottomrule
  \end{tabu}%
\end{table*}

\section{Study}\label{sec:study}

This work aims to assess the impact of improving user and stakeholder confidence by visualizing the stopword removal method and supporting the extraction of a corpus-specific list. Specifically, we want to answer the following four research questions:

\begin{itemize}
\itemindent=2pt
\labelsep=3pt
    \item [\RQOn] Does the method (GPC model) provide a correct evaluation of a word to be a stopword and lead to a stopword list that avoids over-removal while catching corpus-specific stopwords?
    \item [\RQTw] Does a 2-D visualization (GPC Matrix) increase the interpretability of the model?
    \item [\RQTh] Does an interactive visualization interface help users make well-founded decisions?
    \item [\RQFo] How does the approach impact confidence in removing stopwords?
\end{itemize}

To answer our research questions, we designed a four-phase deductive qualitative study. We use a representative academic corpus of university research publications from many corpora for our data and stimuli since it allows us to make more accurate assumptions and judgements based on our experience. Twelve participants at least at the postdoctoral level in the Mathematics and Computer Science departments were recruited to complete the study. Their working experience in the academic areas, familiarity with the publications, and understanding of the corpus make them a representative group of \textbf{expert users}, able to explore the topic model.

The study uses semi-structured interviews. Each interview session was supported by a task questionnaire during which participants were guided through the appropriate tasks. They were encouraged to ask questions and raise points for discussion during the session. In the first four participant sessions, the discussion was fully integrated with the task questionnaire, while the last eight participant sessions ended with a formal discussion after the questionnaire was completed. Each full session was recorded in video, and transcriptions were made from the audio tracks. The study consists of four phases. During each phase, participants are asked to interact with the interface to complete tasks and answer questions based on their judgment. The questionnaire contains four kinds of questions: open, closed, Likert scale, and semantic differential items. In the study, we define four stopword lists as described in Table~\ref{tab:lists}. The discussion session is about their experience with the 2-D visualization. The semi-structured interview format allowed us to explain the visualization features and probe participants' experience using the visualization.



During \textbf{Phase 1 (initial phase)}, we posed an initial scenario to introduce the stopword removal tasks and help participants be aware of the consequences caused by wrongly removed stopwords. At this stage, the participants were set to be in charge of the university's research department preparing an annual academic review based on publications. They were informed how the visualized topic model could help and how stopword removal could influence the visualization outcome. 

We provided two guidelines to our participants: (a) make sure not to remove any words worth generating the topic, and (b) eliminate meaningless words as much as possible. These are relevant to \RQOn~and our motivation for choosing GPC. In this phase, participants needed to choose from L1 and L2. We asked their intuitive opinion about how removing these lists can influence the corpus.

Then, for \textbf{Phase 2 (interaction phase)}, we provided the participants with the evaluation of a word being a stopword given by the GPC model and the visualization interface. In this phase, we provided a list of word and $p_t$ pairs (shown in Table \ref{tab:pt}). The participants were asked how well these $p_t$s fit their intuition (\RQOn~correctness) and how confident they felt (\RQFo) using them.

\begin{table}
  \caption{Manually selected word-$p_t$ pairs presented to participants, used to compare their intuition against GPC metrics.}
  \label{tab:pt}
  \begin{center}
  \begin{tabular}{c c|c c|c c}
    \textbf{word} & $p_t$  & \textbf{word} & $p_t$& \textbf{word} & $p_t$\\ \hline
    this &	$12.56\%$ & 
    study &	$38.98\%$ & 
    not	& $43.13\%$ \\
    paper &	$53.15\%$ & 
    carbon &	$64.63\%$ & 
    complex &	$65.62\%$ \\ 
    structure &	$66.43\% $& 
    robot &	$71.32\%$ & 
    bacterium &	$71.63\%$\\ 
    oil &	$71.78\%$ \\ 
\end{tabular}
\end{center}
\end{table}

To evaluate \RQTh~(decision support) we tasked participants to operate the threshold selector in the interface and decide on an ideal threshold. We recorded the threshold participants chose and asked whether there were any missing, miss-classified or unsure words under the threshold. We then assessed the comprehensibility of the threshold selection and the cause of incomprehension. 
 
The phase then focused on \RQOn~(specifically addressing the performance of the stop word extraction) and \RQFo~(confidence in the extraction without extra explanation). To do this L3 was revealed and participants were asked how confident they were in the suitability of L3. Participants were asked to choose from L1 or L2+L3 to evaluate whether the GP-extracted stopwords were as comprehensive as the long traditional list.

In the final step in this phase, to further probe the impact of the extracted stopword list on the visual output (\RQOn), participants were shown two topic maps: one generated after extracting L1 and one after extracting L2+L3. They were asked to choose the better topic map and to explain their reasoning. This step allows the comparison of topic models generated by extracting stopwords manually or by applying our extraction method. To avoid bias, we started a new page for this task and swapped the order of the topic maps from the previous tasks.

\textbf{Phase 3 (GPC Matrix phase)} concerned \RQTw. While interacting with the interface and examining the GPC Matrix, participants were asked to rate their understanding of it, whether they cared about the principle of GP and whether the GPC Matrix helped them understand the model. We asked our participants to choose between L4 and L3, and probed their opinion about GP and whether they would use it. Lastly, we asked about any information they felt they needed which was missing from the visualization.

\textbf{Phase 4 (exit phase)} concluded our study, with questions prompting participants to express their overall impression of our method and any insights they had about the whole stopword analysis process.

\section{Results, Insights and Discussion}\label{sec:results}

Twelve experts with postdoctoral practice in the academic area and with an understanding of the corpus participated in the study. Participants completed the four-phase study including the semi-structured interview session and task-based questionnaire in 35 to 50 minutes. No participant reported having visual impairments (including color blindness). Audio transcriptions and answers to the open-text responses from the questionnaire were analysed using NVivo \cite{nvivo} and axial coding \cite{strauss1990basics}. Hierarchical nodes allowed categorisation of the views and concepts expressed by the participants, coding also provided structure to the results and facilitated quantifying the different views amongst the participants. From the results of the study, 10/12 participants expressed new ideas and extended thoughts about the approach, while 8/12 participants expressed useful insights and comments on the GPC method during the exit phase. Analysis of the results and participant numbers indicate saturation in the coding which enabled us to comprehensibly answer the proposed research questions and provide ample insights for the discussion of the stopword exploration.

\subsection{Understanding Stopword List Selection (\RQOn, \RQFo)}

We tracked participants’ chosen stopword lists through the study phases to explore their selection and how our method affects them when removing stopwords. In Phase 1 participants did not show a united selection (split: 6,6) between choosing L1 (manual curated) or L2 (training practitioner agreed) stopword lists. This result indicates that \textbf{participants could not accurately judge whether a stopword list might over-operate in a corpus only based on a list of some stop words}.

In Phase 2 (threshold list exploration), most of the participants (9/12) thought that the probabilities were helpful, while the rest required more explanation. No participants believe that the threshold was different from their expectations, this indicates that, in general, \textbf{the probabilities fit participants’ mental judgements well}. Furthermore, the majority of the participants (seven) expressed positive partial confidence in the probabilities, three found it difficult to determine confidence, two felt partially unconfident, but no participants completely trusted the probabilities without further explanation. This indicates that \textbf{probabilities are useful but they need to be complemented to fully gain participants’ confidence in them}.

\begin{table}
  \caption{Table of all participants' chosen threshold, and comparison between the correct extension and the confident group with the partially correct extension and the uncertain group}
  \label{tab:threshold}
  \begin{center}
  \begin{tabular}{c|c|c|c|c}
    \toprule
    Threshold & \multicolumn{4}{c}{Participant judgement} \\
    selected & \textit{correct} & \textit{part correct} & \textit{confident} & \textit{uncertain} \\
    \hline

    58.00 & & \checkmark & & \checkmark \\
59.93 & & \checkmark & & \checkmark \\
61.00 & & \checkmark & & \checkmark \\
\hline
63.87 &\checkmark&   & & \checkmark \\
64.00 & & \checkmark &\checkmark&   \\
64.16  &\checkmark&   &\checkmark& \\
\hline
66.40 &\checkmark&   &\checkmark&  \\
66.87 &\checkmark&   &\checkmark&  \\
66.93 &\checkmark&   &\checkmark&  \\
67.00 & & \checkmark & & \checkmark \\
67.10 &\checkmark&   & \checkmark & \\
68.00 &\checkmark&   &\checkmark&  \\
\bottomrule

\end{tabular}
\end{center}
\end{table}


Table~\ref{tab:threshold} shows the participants’ chosen thresholds. 9/12 participants’ thresholds lay in the range of 64-68. In comparison, three participants selected the range between 58-61, which significantly differs from the typical selection range. These indicated that some experts were very cautious about the words to be removed. In this regard, this suggests that \textbf{the threshold selection allowed them to decide based on their level of caution}. This is interesting as caution receives little consideration by practitioners, but it important for decision-making and confidence in output visualizations.

Phase 2, L3 (GPC corpus-specific <60\% list) as an extension of L2 (training practitioner list) was also explored. Seven participants agreed that the extension was all correct, while the rest (five) considered it partially correct. The participants who trusted the model result were more conservative when removing words. When asked whether they felt confident about the extension, half of the participants (six) mentioned that they felt confident, while the other half could not tell. The comparison of the selected thresholds between the correct extension, the confident, the partially correct and the uncertain group is shown in Table ~\ref{tab:threshold}. These results reinforce a possible common deduction that \textbf{cautious users who remove fewer words have lower confidence in the model result}.

During the study, out of the 12 participants, six participants originally chose the automated list (L2 + L3) and six chose the manually curated list (L1). However, three participants changed their choice from (L1) to (L2 + L3) explaining that several words they were expert in were missed in (L1). Two participants changed from (L2 + L3) to (L1). P2, for example, thought there were words in the automated list that described research topics and therefore shouldn't be removed. P9 thought (L1) it was more comprehensive. This shows that \textbf{factors such as the experience of the practitioner and the number of stopwords heavily influence the selection}.

Looking at the visualisation outputs using the different stopword lists, out of the 12 participants, seven chose the map visualization generated by the automated extended list (L2 + L3), and the rest chose the other (L1). The number of participants who prefered the automated extended list (L2 + L3) is equal in number to the participants choosing the maps generated using this list, albeit these are not matches, six participants chose a generated map not using their preferred list. Interestingly, users may not get a better result after carefully selecting a list of stopwords. As so, we can infer that \textbf{the automated extended list by our GPC pipeline can provide perceptually the same level of stopword removal as an manually curated long list}.

In Phase 3, we used the approximate GPC Matrix to extract a stopword list (L4) under the same conditions as the 30-D GP model. We let our participants choose from these two lists. From the 12 experts, nine participants chose the 30-D version list, indicating that the 2-D model decreased the performance of extracting stopwords. Also, from the 12 participants, five would like to consider the GPC Matrix as an approximation that helps them understand the model; despite there being a loss in the performance, while the five other participants thought that the 2-D visualization should also be used at stopword extraction since it can be accurately visualized. The remaining two participants thought that was not needed. As a result, we can imply that \textbf{most would still prefer the aid of the GPC Matrix visualization}.

All participants (12) preferred to have control of the removal of stopwords, and all agreed that this \textbf{control increases their confidence}. Most participants (7) would like to use the tool \textbf{to support them in the removal and estimation of stopwords}; the rest (5) would consider using it \textbf{in high-consequence scenarios}. These results indicate a clear need for a better stopword approach in both cases.

\subsection{Exploring the Threshold Selection, Interface and Visualizations (\RQOn, \RQTw, \RQTh)}

During Phase 2 of the study, participants were asked to complete tasks based on the interface (Figure 1). All participants (12) \textbf{preferred control over stopword selection}, from all, nine of the participants agreed that threshold selection is easy to use and also nine understood well how the slider divides words by setting a threshold.

We avoided introducing over-complicated mathematical concepts in the interface as this can lead to positive bias (e.g. users tend to trust complex mathematics rather than review the correctness). However, \textbf{participants who cared about decision-making wished to know more about the theory behind the interface}.

Exploring the interface, we found three significant concerns in our participants: (a) P3 and P10 found it hard to get a clear idea of how the algorithm works. \textbf{They required more explanation to link things together}. For example, P3 mentioned ``\textit{I thought that the thing (method) behind the interface was also useful (to know)''}; (b) participants found some words included that needed to be corrected (8/12), while some words missed the threshold (10/12), and some words were difficult to judge (8/12). For example, P11 stated that ``\textit{it is currently a "one size fits all" system; need to tweak specific words near the threshold.}'' We include the threshold selection feature to avoid making repeat choices. However, the participants have different ideas about whether a word should be removed, which indicates that \textbf{there is no perfect list of stopwords. Therefore, instead of being a selector, the threshold could be used as an advisor}, this also allows further involvement and ownership from participants of the outputs; finally, (c) P10 noted that the interface only shows the details of a single topic at a time, making comparisons over topics difficult. P8 suggested that having a list that contains all the words extracted by the threshold would help. Stopword selection is a process in which users need information throughout all topics in the corpus instead of within one to make a confident decision. We also observed that the visualization hierarchy of the topic occasionally increases the difficulty in understanding the information. Moreover, \textbf{comprehensive decisions require a general picture of the whole} (for example, when determining the thresholds).

From the analysis and some individual results, we can observe \textbf{three fundamental interface design insights} that practitioners can implement in their own work: (a) users typing a word in a familiar area can increase confidence as noted by P12. Selecting several words and seeing how their probabilities are distributed can help users determine the threshold as noted by P10. Since it is hard to grasp the overall situation of the whole corpus, starting from \textbf{familiar terms} allows users to understand the threshold and test the model’s robustness. Thus, adding a \textbf{bi-directional exploration} (browsing <=> search) as a feature could increase confidence when making comprehensive decisions about a large corpus; (b) an animated sequential interaction, like a play button, instead of a slider could help \textbf{reduce mental load} and improve users’ understanding of how the stopwords incrementally change rather than having to change, look, and think simultaneously as noted by P12; finally, (c) P7 wished to differentiate if an extracted word is a universal stopword or a corpus-specific one. Users could have different standards for these two types of stopwords. \textbf{Contrasting} universal and corpus-specific stopwords would help users make better judgements.

Furthermore, we observed insights that can be extrapolated into \textbf{four essential recommendations}: (a) for removal tasks (e.g. stopword), where users estimate terms, displaying an \textbf{initial list} can provide an intuitive starting result for users; (b) the selection can be initially set in a \textbf{preliminary ideal} or automatic setting. This will generally satisfy their requirements, but users will still often play around with the threshold in different ways. This interaction or human adjustment can greatly improve user confidence in the output and result; (c) Users tend to use \textbf{judgements based on familiar areas}. Starting with familiar words (e.g., allow bi-directional interaction) could help users build confidence and support them in making better choices; and (d) when designing an interface that has interactivity for debatable matters (such as stopword removal), it is important to be \textbf{aware of the different requirement} levels to support making decisions. Letting the user decide on the flexibility of the interface could be a valuable resource to increase the user’s confidence in such situations.

\subsection{Analyzing the GPC Matrix (\RQTh)}

As previously discussed, we observed that there is a barrier for non-specialist users to understand the 2-D visualization. Most participants (11/12) needed an extra explanation to understand the principle of 2-D visualization. However, a participant with a strong mathematics background, P11, completely understood it. When asked whether the visualisation increased the willingness to use GPC, P11 was the only one to give a positive answer. Seven participants thought the 2-D visualization did not help them understand the model. Participants (eight) discussed this issue further in Phase 4 (exit phase) as an open discussion, a brief explanation of what a \textit{dimension} means, how it relates to a topic, and why we made a 2-D approximation of the GP model resulted in all eight participants it was noted that they managed to fully understand the principle of the 2-D Visualization. These observations and the views expressed by the participants indicate that \textbf{the participants' understanding of the visualization is essential for deciding whether it is a valuable addition}.

Users had two main concerns about understanding the visualisation. Firstly, participants perceived a disconnect between the 2-D visualization panel and the other parts of the interface. For example, P12 expressed:``\textit{...it is unclear how it links to other elements on the page}''. Also, P1 and P3 stated that the background color of GPC Matrix did not change with the slider. The Matrix shows the trained model given by the 2-D Gaussian process model. Modifying the threshold will not trigger a change. Adding change driven by elements in the threshold can avoid these disconnection issues. Secondly, P1, P5, P6 and P11 could not understand what dimension stands for, and what it is doing with the frequency. P8 asked for the reason for all the distributions to decrease monotonically. Because of the pre-processing that we did in Section \ref{sec:set}, for rigour consideration, we call the x-axis of GPC Matrix a dimension instead of a topic, which leads to confusion for users. However, this was mitigated as participants could understand the reason with an additional explanation. Therefore, although the \textbf{GPC Matrix can facilitate the exploration of the stopwords and selection, it is still recommended to add an explanation of the pre-process for clarification} to improve the link between the model and visualization, and to an extent, this explanation will also improve confidence in users.

\subsection{Exploring the Method and Principle (\RQOn, \RQTw, \RQTh, \RQFo)}

Of all the participants (12) in our study, nine cared about the principle behind the method. \textbf{Knowing the principle increased their confidence in using it}. Participants mentioned that understanding the principle replaces their need to review every choice. Knowing how the topic map was constructed allowed them to feel informed and let them be cautious when generating the topic map. On the contrary, the other three participants were concerned about accuracy rather than the principle. P6, who described himself as a result-oriented individual, mentioned that \textbf{they would use an approach only if it proved accurate}.

We noted two concerns from our participants about our method: (a) P12 mentioned that context should be considered when evaluating stopwords. Current frequency calculation might need to be revised because the meaning of words changes throughout the different contexts. \textbf{This issue arises when users' experience of specific phrases does not match the model result; this can be solved by human intervention or by combining the approach with encoder-based context-sensitive approaches} such as BERT \cite{bert}; (b) P9 raised a concern about considering the $Swdf$ as a normal distribution since Gaussian processes require it. The participant had a robust math background, so we considered this an expert area of concern. We would note that the $Swdf$ fits a multivariate Gaussian distribution, which is an assumption made since we considered the use of words by writers in each topic to follow a normal distribution. It is important to note that the participants' concerns are a welcomed reassurance to their increased confidence in knowing and understanding the principle behind the approach.

\subsection{Summary}

In short, we found that stopword removal is not a quintessentially simple mechanical task, for example, participants struggled to judge some aspects of stopword lists, while factors affected their understanding and confidence, and our exploration provided many interesting insights on this essential task.

To begin, we showed that \textbf{our method and stopword probabilities fit well with the mental models or expectations of our participants} (\RQOn). These results show that the studied methods and probabilities can work, but to fully gain confidence in participants we recommend these are complementary with simple explanations. Threshold selections from our method can be useful to improve confidence in participants, moreover, these can be linked to the level of caution of the participants which is important for confidence in various tasks, especially ones involving decisions and stakeholders. Finally, factors including experience, caution and perceived length of lists can influence the selection of stopwords and need to be considered.

We found our method can provide the same perceptual level of stopword removal as a time-consuming lengthy manually curated stopword list. As expected, the stopword list from the approximate GPC Matrix abstraction was considered simpler but not preferred to the complete GPC Matrix list, but most participants still preferred to have the aid of the GPC Matrix visualization. 

One important aspect of the exploration was that all \textbf{participants preferred control over the removal of stopwords and we observed that control increased their confidence} (\RQFo). Participants expressed the need for further support in the estimation and removal, especially for high-consequence scenarios where confidence in the output is a necessity.
When exploring the threshold selection and interface, we found that control over the selection was preferred. Moreover, there was an appetite to know more about the interface workings and this was more prevalent for high-risk and decision-making tasks. We found that there is no perfect stopword list and the threshold selection can be used as an advisor mechanism to improve understanding and confidence. 

\textbf{Insights seen in the exploration can aid practitioners} (\RQTh). These include fundamental interface recommendations like using familiar terms, bi-directional exploration, reduction of mental load, and contrasting terms. Furthermore, essential recommendations can be added as having an initial list or output, setting preliminary settings, having familiar areas and an awareness of different requirements. 

\textbf{The GPC Matrix helped facilitate the exploration and selection of stopwords for participants} (\RQTw), but their understanding of the visualization is essential for deciding whether this would be a valuable addition to an interface. We found that an explanation of the pre-process for the visualization for clarification is recommended. This also shows that participants not only think about the interface, but they also reflect on the origin or provenance of the information throughout it. So, when solely removing the standard set of stopwords to models to create output visualisations, practitioners should consider the ramifications of doing so and how this will impact users understanding, perception and confidence.

Finally, exploring the method, it was reasserted that knowing the underlying principles increased confidence in using the interface and visualization outputs. We also found, however, that perceived accuracy as well plays a significant role. Therefore, participants not only need to know the methods or principles but they also need to perceive these as accurate to improve confidence.

\subsection{Limitations and Future Work}

In this section, we will discuss the limitations of our work and the opportunities for future work.

Firstly, as with any dimensionality reduction method, the GPC model drops in performance as its number of dimensions increases. However, the 2-D approximation principle presented in this paper shows one working solution to this problem.

In an attempt to reduce complexity, we abstracted stopword removal to a single threshold selection. While simple to use, this one-size-fits-all approach is not ideal for all scenarios. This can be addressed in future investigations, where additional interaction could further tweak controversial stopwords.

Our study primarily targeted expert practitioners, with knowledge of the corpus. While their opinions are valuable and provide deep insights into confident stopword removal, we recognize that a more varied audience would first need to build their mental model of the corpus before using our approach and visualization.

Our method can be integrated into existing topic modelling systems in future work, both automated and interactive. For instance, the GPC modelling and probability estimation can be added to pipelines for automatic corpus cleaning. As well as the computation and display of probabilities can open new avenues for improved visualization and annotation of topic models.

\section{Conclusion}

In this work, we aim to improve the use of visualizations in Machine Learning outputs, in particular with the analysis of stopword removal in text corpora. While several works have proposed computational scores or metrics, we introduce a combined probabilistic and visual solution to facilitate the exploration of stopword removal and evaluate how it influences confidence. 

We applied Gaussian processes classification (GPC) to model latent stopword distributions in a corpus, using a limited set of labelled input. With this novel method, we were able to extract a list of stopwords -- and their probabilities -- consisting of both common and corpus-specific terms.
We integrated these results within a topic model visualization system by (a) providing an interactive threshold selector to highlight extracted stopwords and (b) developing a new visualization module (the GPC Matrix) to illustrate the GPC model to practitioners.
We applied this model, threshold selector and GPC Matrix in a study to explore how it influences practice and user confidence in stopword analysis, including the impact of their removal on topic models, visualizations and its subsequent effects (e.g., decision-making tasks).

Our results show that the GPC-based stopword extraction method coincided with participants' mental models and intuitions about stopwords. Furthermore, the interaction threshold analysis strongly improved their confidence. Finally, despite the complexity of the GPC Matrix, with clarifications, most participants agreed that it helped facilitate their exploration of stopword probabilities.

Along with these results, we discuss insights and recommendations for practitioners to improve the visual exploration of stopwords. These include suggested practices to support stakeholders with similar Machine Learning outputs.


\acknowledgments{%
The authors would like to thank the participants for their involvement in this study and their constructive insights.
This work was supported by [anonymized].
}

\bibliographystyle{IEEEtran}

\bibliography{template}

\begin{thebibliography}{10}
\providecommand{\url}[1]{#1}
\csname url@samestyle\endcsname
\providecommand{\newblock}{\relax}
\providecommand{\bibinfo}[2]{#2}
\providecommand{\BIBentrySTDinterwordspacing}{\spaceskip=0pt\relax}
\providecommand{\BIBentryALTinterwordstretchfactor}{4}
\providecommand{\BIBentryALTinterwordspacing}{\spaceskip=\fontdimen2\font plus
\BIBentryALTinterwordstretchfactor\fontdimen3\font minus
  \fontdimen4\font\relax}
\providecommand{\BIBforeignlanguage}[2]{{%
\expandafter\ifx\csname l@#1\endcsname\relax
\typeout{** WARNING: IEEEtran.bst: No hyphenation pattern has been}%
\typeout{** loaded for the language `#1'. Using the pattern for}%
\typeout{** the default language instead.}%
\else
\language=\csname l@#1\endcsname
\fi
#2}}
\providecommand{\BIBdecl}{\relax}
\BIBdecl

\bibitem{le2020visualising}
P.~Le~Bras, A.~Gharavi, D.~A. Robb, A.~F. Vidal, S.~Padilla, and M.~J.
  Chantler, ``Visualising covid-19 research,'' \emph{arXiv preprint
  arXiv:2005.06380}, vol.~1, 2020.

\bibitem{stopwordRethinking}
A.~Schofield, M.~Magnusson, and D.~Mimno, ``Pulling out the stops: Rethinking
  stopword removal for topic models,'' in \emph{Proceedings of the 15th
  Conference of the European Chapter of the Association for Computational
  Linguistics: Volume 2, short papers}, 2017, pp. 432--436.

\bibitem{2020stopwordIr}
D.~J. Ladani and N.~P. Desai, ``Stopword identification and removal techniques
  on tc and ir applications: A survey,'' in \emph{2020 6th International
  Conference on Advanced Computing and Communication Systems (ICACCS)}.\hskip
  1em plus 0.5em minus 0.4em\relax IEEE, 2020, pp. 466--472.

\bibitem{boyd2014care}
J.~Boyd-Graber, D.~Mimno, and D.~Newman, ``Care and feeding of topic models:
  Problems, diagnostics, and improvements,'' \emph{Handbook of mixed membership
  models and their applications}, vol. 225255, 2014.

\bibitem{corpusSpecStopwords}
V.~P. Baradad and A.-m. Mugabushaka, ``Corpus specific stop words to improve
  the textual analysis in scientometrics.'' in \emph{ISSI}, 2015.

\bibitem{wang-etal-2020-cord}
\BIBentryALTinterwordspacing
L.~L. Wang, K.~Lo, Y.~Chandrasekhar, R.~Reas, J.~Yang, D.~Burdick, D.~Eide,
  K.~Funk, Y.~Katsis, R.~M. Kinney, Y.~Li, Z.~Liu, W.~Merrill, P.~Mooney, D.~A.
  Murdick, D.~Rishi, J.~Sheehan, Z.~Shen, B.~Stilson, A.~D. Wade, K.~Wang,
  N.~X.~R. Wang, C.~Wilhelm, B.~Xie, D.~M. Raymond, D.~S. Weld, O.~Etzioni, and
  S.~Kohlmeier, ``{CORD-19}: The {COVID-19} open research dataset,'' in
  \emph{Proceedings of the 1st Workshop on {NLP} for {COVID-19} at {ACL}
  2020}.\hskip 1em plus 0.5em minus 0.4em\relax Online: Association for
  Computational Linguistics, Jul. 2020. [Online]. Available:
  \url{https://www.aclweb.org/anthology/2020.nlpcovid19-acl.1}
\BIBentrySTDinterwordspacing

\bibitem{sievert2014ldavis}
C.~Sievert and K.~Shirley, ``Ldavis: A method for visualizing and interpreting
  topics,'' in \emph{Proceedings of the workshop on interactive language
  learning, visualization, and interfaces}, 2014, pp. 63--70.

\bibitem{bert}
J.~Devlin, M.-W. Chang, K.~Lee, and K.~Toutanova, ``Bert: Pre-training of deep
  bidirectional transformers for language understanding,'' \emph{arXiv preprint
  arXiv:1810.04805}, 2018.

\bibitem{Churchill2022}
\BIBentryALTinterwordspacing
R.~Churchill and L.~Singh, ``The evolution of topic modeling,'' \emph{ACM
  Comput. Surv.}, vol.~54, no. 10s, nov 2022. [Online]. Available:
  \url{https://doi.org/10.1145/3507900}
\BIBentrySTDinterwordspacing

\bibitem{manning2014stanford}
C.~D. Manning, M.~Surdeanu, J.~Bauer, J.~R. Finkel, S.~Bethard, and
  D.~McClosky, ``The stanford corenlp natural language processing toolkit,'' in
  \emph{Proceedings of 52nd annual meeting of the association for computational
  linguistics: system demonstrations}, 2014, pp. 55--60.

\bibitem{grootendorst2022bertopic}
M.~Grootendorst, ``Bertopic: Neural topic modeling with a class-based tf-idf
  procedure,'' \emph{arXiv preprint arXiv:2203.05794}, 2022.

\bibitem{nothman2018stop}
J.~Nothman, H.~Qin, and R.~Yurchak, ``Stop word lists in free open-source
  software packages,'' in \emph{Proceedings of Workshop for NLP Open Source
  Software (NLP-OSS)}, 2018, pp. 7--12.

\bibitem{churchill2018}
R.~Churchill, L.~Singh, and C.~Kirov, ``A temporal topic model for noisy
  mediums,'' in \emph{Advances in Knowledge Discovery and Data Mining},
  D.~Phung, V.~S. Tseng, G.~I. Webb, B.~Ho, M.~Ganji, and L.~Rashidi,
  Eds.\hskip 1em plus 0.5em minus 0.4em\relax Cham: Springer International
  Publishing, 2018, pp. 42--53.

\bibitem{gerlach2019universal}
M.~Gerlach, H.~Shi, and L.~A.~N. Amaral, ``A universal information theoretic
  approach to the identification of stopwords,'' \emph{Nature Machine
  Intelligence}, vol.~1, no.~12, pp. 606--612, 2019.

\bibitem{hao2008automatic}
L.~Hao and L.~Hao, ``Automatic identification of stop words in chinese text
  classification,'' in \emph{2008 International conference on computer science
  and software engineering}, vol.~1.\hskip 1em plus 0.5em minus 0.4em\relax
  IEEE, 2008, pp. 718--722.

\bibitem{joachims1997probabilistic}
T.~Joachims \emph{et~al.}, ``A probabilistic analysis of the rocchio algorithm
  with tfidf for text categorization,'' in \emph{ICML}, vol.~97.\hskip 1em plus
  0.5em minus 0.4em\relax Citeseer, 1997, pp. 143--151.

\bibitem{chuang2012termite}
J.~Chuang, C.~D. Manning, and J.~Heer, ``Termite: Visualization techniques for
  assessing textual topic models,'' in \emph{Proceedings of the international
  working conference on advanced visual interfaces}, 2012, pp. 74--77.

\bibitem{taddy2012}
\BIBentryALTinterwordspacing
M.~Taddy, ``On estimation and selection for topic models,'' in
  \emph{Proceedings of the Fifteenth International Conference on Artificial
  Intelligence and Statistics}, ser. Proceedings of Machine Learning Research,
  N.~D. Lawrence and M.~Girolami, Eds., vol.~22.\hskip 1em plus 0.5em minus
  0.4em\relax La Palma, Canary Islands: PMLR, 21--23 Apr 2012, pp. 1184--1193.
  [Online]. Available: \url{https://proceedings.mlr.press/v22/taddy12.html}
\BIBentrySTDinterwordspacing

\bibitem{wallach2009rethinking}
H.~M. Wallach, D.~M. Mimno, and A.~McCallum, ``Rethinking lda: Why priors
  matter,'' in \emph{Advances in neural information processing systems}, 2009,
  pp. 1973--1981.

\bibitem{sarica2021stopwords}
S.~Sarica and J.~Luo, ``Stopwords in technical language processing,''
  \emph{Plos one}, vol.~16, no.~8, p. e0254937, 2021.

\bibitem{choi2018}
\BIBentryALTinterwordspacing
M.~Choi, S.~Shin, J.~Choi, S.~Langevin, C.~Bethune, P.~Horne, N.~Kronenfeld,
  R.~Kannan, B.~Drake, H.~Park, and J.~Choo, ``Topicontiles: Tile-based
  spatio-temporal event analytics via exclusive topic modeling on social
  media,'' in \emph{Proceedings of the 2018 CHI Conference on Human Factors in
  Computing Systems}, ser. CHI '18.\hskip 1em plus 0.5em minus 0.4em\relax New
  York, NY, USA: Association for Computing Machinery, 2018, p. 1–11.
  [Online]. Available: \url{https://doi.org/10.1145/3173574.3174157}
\BIBentrySTDinterwordspacing

\bibitem{liu2020mapping}
H.~Liu, Z.~Chen, J.~Tang, Y.~Zhou, and S.~Liu, ``Mapping the technology
  evolution path: a novel model for dynamic topic detection and tracking,''
  \emph{Scientometrics}, vol. 125, pp. 2043--2090, 2020.

\bibitem{abdul2018}
\BIBentryALTinterwordspacing
A.~Abdul, J.~Vermeulen, D.~Wang, B.~Y. Lim, and M.~Kankanhalli, ``Trends and
  trajectories for explainable, accountable and intelligible systems: An hci
  research agenda,'' in \emph{Proceedings of the 2018 CHI Conference on Human
  Factors in Computing Systems}, ser. CHI '18.\hskip 1em plus 0.5em minus
  0.4em\relax New York, NY, USA: Association for Computing Machinery, 2018, p.
  1–18. [Online]. Available: \url{https://doi.org/10.1145/3173574.3174156}
\BIBentrySTDinterwordspacing

\bibitem{chaney2012visualizing}
A.~Chaney and D.~Blei, ``Visualizing topic models,'' in \emph{Proceedings of
  the International AAAI Conference on Web and Social Media}, vol.~6, no.~1,
  2012, pp. 419--422.

\bibitem{doshi2017towards}
F.~Doshi-Velez and B.~Kim, ``Towards a rigorous science of interpretable
  machine learning,'' \emph{arXiv preprint arXiv:1702.08608}, 2017.

\bibitem{roscher2020explainable}
R.~Roscher, B.~Bohn, M.~F. Duarte, and J.~Garcke, ``Explainable machine
  learning for scientific insights and discoveries,'' \emph{Ieee Access},
  vol.~8, pp. 42\,200--42\,216, 2020.

\bibitem{rasmussen2004gaussian}
C.~E. Rasmussen, \emph{Gaussian processes in machine learning}.\hskip 1em plus
  0.5em minus 0.4em\relax Springer, 2004.

\bibitem{lda}
D.~M. Blei, A.~Y. Ng, and M.~I. Jordan, ``Latent dirichlet allocation,''
  \emph{Journal of machine Learning research}, vol.~3, no. Jan, pp. 993--1022,
  2003.

\bibitem{Reuters-RCV1}
M.~R. Amini, N.~Usunier, and C.~Goutte, ``Learning from multiple partially
  observed views-an application to multilingual text categorization,''
  \emph{Advances in neural information processing systems}, vol.~22, 2009.

\bibitem{scikit-learn}
F.~Pedregosa, G.~Varoquaux, A.~Gramfort, V.~Michel, B.~Thirion, O.~Grisel,
  M.~Blondel, P.~Prettenhofer, R.~Weiss, V.~Dubourg, J.~Vanderplas, A.~Passos,
  D.~Cournapeau, M.~Brucher, M.~Perrot, and E.~Duchesnay, ``Scikit-learn:
  Machine learning in {P}ython,'' \emph{Journal of Machine Learning Research},
  vol.~12, pp. 2825--2830, 2011.

\bibitem{Shneiderman2003}
\BIBentryALTinterwordspacing
B.~Shneiderman, ``The eyes have it: A task by data type taxonomy for
  information visualizations,'' in \emph{The Craft of Information
  Visualization}, ser. Interactive Technologies, B.~B. BEDERSON and
  B.~SHNEIDERMAN, Eds.\hskip 1em plus 0.5em minus 0.4em\relax San Francisco:
  Morgan Kaufmann, 2003, pp. 364--371. [Online]. Available:
  \url{https://www.sciencedirect.com/science/article/pii/B9781558609150500469}
\BIBentrySTDinterwordspacing

\bibitem{nvivo}
K.~Jackson and P.~Bazeley, \emph{Qualitative data analysis with NVivo}.\hskip
  1em plus 0.5em minus 0.4em\relax Sage, 2019.

\bibitem{strauss1990basics}
A.~Strauss and J.~Corbin, \emph{Basics of qualitative research}.\hskip 1em plus
  0.5em minus 0.4em\relax Sage publications, 1990.

\end{thebibliography}

\appendix 

\section{Pilot Experiments}\label{app:pilot}

This appendix describes our pilot experiments and observations to evaluate the robustness of the approach.

\subsection{The comparison of using the set of document frequency and term frequency}
To find an ideal input feature that provides better performance, we compared the outcomes using the set of document frequency, $\mathbf{Swdf}_j$, with results using the set of term frequency, $\mathbf{Swtf}_j$. Similarly as $\mathbf{Swdf}_j$, we define 
$\mathbf{Swtf}_j=\{Tfw_{j,i}\}_{i=1}^{30}$,
where $Tfw_{j,i}$ is the frequency of word $w_j$ over topic $i$. $N^{(i)}_w$ is the number of total words in the $i^{th}$ topic, $Nw_{j,i}$ is how many times the word $w_j$ appears in topic $i$. Thus $Tfw_{j,i}$ can be calculated as:
\begin{equation}
     Tfw_{j,i} = \frac{Nw_{j,i}}{N^{(i)}_w}\qquad \mathrm{for}\; i = 1,\;2,\cdots,\;30.
\end{equation}
We chose to compare document frequency to term frequency because it is a typical benchmark for collecting stopwords. After comparing $\mathbf{Swtf}_j$ with $\mathbf{Swdf}_j$, document frequency showed a stronger classification capability. We list two reasons below to explain why document frequency is a better choice.

We first notice that $\mathbf{Swdf}_j$ has a better description of how selected words distribute in different topics, especially in extreme cases. Given two topics ($i=1,\;2$), which both have \textbf{ND} documents and \textbf{NW} words in each document. The total number of words in each topic will be $N^{(1)}_w =N^{(2)}_w =\mathbf{ND}\times \mathbf{NW}$. For a particular word $w_{\ast}$, we present two hypothetical situations of how this word distributes in these two topics in table \ref{tab:assume2top}. In topic 1, we assume the word is only in one document but appears $\mathbf{ND}$ times, while in topic 2, appears in every document but only once. As table \ref{tab:assume2top} shows, the term frequency $Tfw_{\ast,i}$ value stays the same in two topics, which means $\mathbf{Swtf}_j$ is not as sensitive as $\mathbf{Swdf}_j$ in these extreme situations.

\begin{table}[h]
  \caption{Two assumed behaviours for word $w_{\ast}$ in two topics}
  \label{tab:assume2top}
  \begin{center}
   \renewcommand{\arraystretch}{1.7}
  \begin{tabular}{c|c|c}
    &\textbf{Topic 1} & \textbf{Topic 2}\\ \hline
     $Ndw_{\ast}$& 1 & \textbf{ND}  \\ \hline
		$w_{\ast}d$ & \textbf{ND} & 1  \\ \hline
		$Nw_{\ast,i}$ & $Nw_{\ast,1}$= $1\times \mathbf{ND}=\mathbf{ND}$ & $Nw_{\ast,2}$= $\mathbf{ND} \times 1=\mathbf{ND}$  \\ \hline
		$Dfw_{\ast,i}$  & $\frac{1}{\mathbf{ND}}$ & $\frac{\mathbf{ND}}{\mathbf{ND}} =1$ \\ \hline
            $Tfw_{\ast,i}$& $ \frac{\mathbf{ND}}{\mathbf{ND}\times\mathbf{NW}}=\frac{1}{\mathbf{NW}}$ & $ \frac{\mathbf{ND}}{\mathbf{ND}\times\mathbf{NW}}=\frac{1}{\mathbf{NW}}$ \\
\end{tabular}
\end{center}
\end{table}

\begin{figure}[h]
  \includegraphics[width=\linewidth]{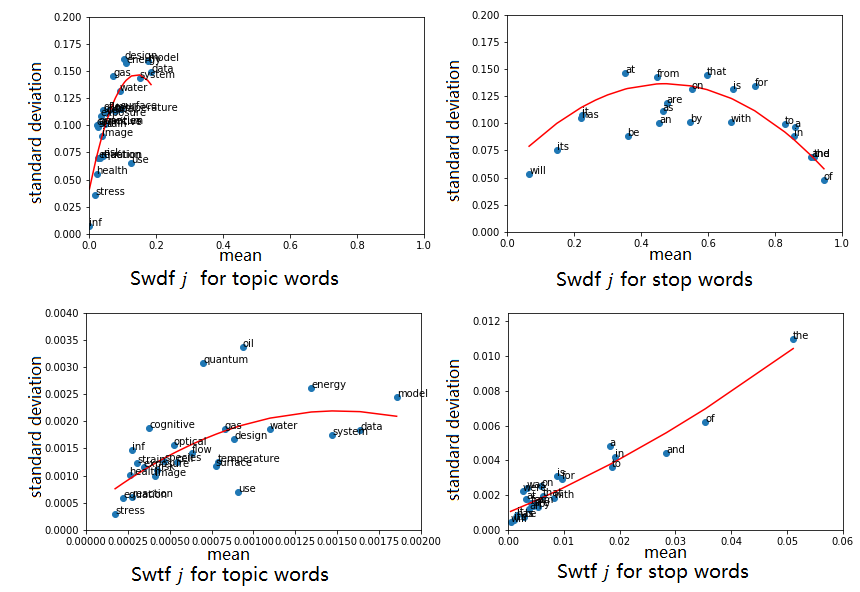}
  \caption{The scatter plots of the mean and standard deviation of the $\mathbf{Swdf}_j$ for topic words (up left), the  $\mathbf{Swdf}_j$ for stopwords (up right), the $\mathbf{Swtf}_j$  for topic words (bottom left) and the $\mathbf{Swtf}_j$  for stopwords (bottom right)}
  \label{fig:dtswtw-joint}
\end{figure}

We then plot four comparison graphs shown as Figure \ref{fig:dtswtw-joint} for the mean and variance of both $\mathbf{Swdf}_j$ and $\mathbf{Swtf}_j$ over 30 topics for selected stop and topic words. Figure \ref{fig:dtswtw-joint} indicate that the layouts of mean and standard deviation of $\mathbf{Swdf}_j$ for topic words are distinct from stopwords. On the contrary, layouts of mean and standard deviation of $\mathbf{Swtf}_j$ for topic words and stopwords are similar. Therefore, we consider that the GPC model can provide better classification results by using $\mathbf{Swdf}$ as the input values.

\subsection{Model performance over different corpora}

We apply the methods to different kinds of corpora, including:

\begin{itemize}
  \item Academic publications in a specific area
  \item Online news in daily language
  \item User questions and AI generated response from chatting system
\end{itemize}
We successfully extracted stopwords for all of the corpora listed above. We did not find any significant decrease in the accuracy of the word evaluation. However, the time consumption of the GPC model is relevant to the corpus size, and the ideal range of threshold to split words changes through corpora.

\subsection{Various choice of kernel function}

This pilot experiment is to determine the kernel function. We applied all suitable pre-defined kernels in \cite{scikit-learn}. The scores of the GPC models generated by these kernel functions are shown in Table \ref{tab:knsc}. The four highest-scoring functions extraction gave almost the same results. By comparing the stopwords extracted by different kernel functions, we found that the Radial-basis function (RBF) is the only one that extracted an additional word: 'from'. Therefore, we consider it has the utmost ability to extract stopwords and choose to use it for this study.

\begin{table}[h]
  \caption{The GPC scores of different kernel functions}
  \label{tab:knsc}
  \begin{center}
  \begin{tabular}{c|c}
\textbf{kernel} & \textbf{score} \\ \hline
Matérn kernel & 0.9167 \\ 
Exp-Sine-Squared kernel & 0.9167 \\ 
Rational quadratic kernel & 0.9167 \\ 
Radial-basis function kernel & 0.9063\\ 
Dot-Product kernel & 0.8958\\ 
White kernel & 0.4688 \\ 
\end{tabular}
\end{center}
\end{table}

\subsection{Changing number of topics}

We change the number of topics in the topic model and evaluate the extraction performance. Since the increase in topic leads to an increase in the dimension of the GPC model, which results in a drop in the performance. The model struggled to provide meaningful classification when the dimension increased beyond $100$. However, we did not observe the performance drop in 2-D approximate version of GPC.

\end{document}